\documentstyle[12pt]{article}
\newcommand{\be}{\begin{equation}}\newcommand{\ee}{\end{equation}}
\newcommand{\bea}{\begin{eqnarray}}\newcommand{\eea}{\end{eqnarray}}
\newcommand{\nn}{\nonumber}\newcommand{\p}[1]{(\ref{#1})}

\begin{document}
\renewcommand{\thefootnote}{\fnsymbol{footnote}}
\thispagestyle{empty}
\begin{center}
{\hfill JINR E2-95-427}\vspace{0.2cm} \\
{\hfill hep-th/9510072}\vspace{2cm} \\
CONFORMAL LINEARIZATION VERSUS NONLINEARITY OF $W$-ALGEBRAS \vspace{1cm} \\
S. Krivonos\footnote{E-mail: krivonos@cv.jinr.dubna.su} and
A. Sorin\footnote{E-mail: sorin@cv.jinr.dubna.su} \vspace{1cm}\\
Bogoliubov Laboratory of Theoretical Physics, JINR,\\
141980, Dubna, Moscow Region, Russia \vspace{3cm} \\
{\bf Abstract}
\end{center}
We review the new approach to the theory of nonlinear $W$-algebras which
is developed recently and called {\it conformal linearization}.
In this approach $W$-algebras are embedded as subalgebras into some
{\it linear conformal} algebras with a finite set of currents and most of
their properties could be understood in a much simpler way by studing
their linear counterpart. The general construction is illustrated by the
examples of $u(N)$-superconformal, $W(sl(N),sl(2))$, $W(sl(N),sl(N))$
as well as $W(sl(N),sl(3))$ algebras. Applications to the construction of
realizations (included modulo null fields realizations) as well as central
charge spectrum for minimal models of nonlinear algebras are discussed.
\vspace{1cm}
\begin{center}
To appear in ``Geometry and Integrable Models'',
Eds.: P.N.Pyatov \& S.N.Solodukhin, \\
World Scientific Publ. Co. (in press)
\end{center}
\vspace{1cm}
\begin{center}
October 1995
\end{center}
\vfill
\setcounter{page}0
\renewcommand{\thefootnote}{\arabic{footnote}}
\setcounter{footnote}0
\newpage

\section{Introduction}

Since the pioneer paper of Zamolodchikov \cite{W3}, a lot of extended
nonlinear conformal algebras (the $W$-type algebras) have been constructed
and studied (see, e.g., \cite{BS} and references therein). The growing
interest to this subject is motivated by many interesting applications of
nonlinear algebras to the string theory, integrable systems, etc.
However, the intrinsic nonlinearity of W-algebras makes it rather difficult
to apply to them the standard arsenal of techniques and means used in the
case of linear algebras (while constructing their field representations,
etc.).

A way to circumvent this difficulty has been proposed by us in
\cite{KS,KSlin1,BKS,KSlin2}. We
found that in many cases a given nonlinear $W$ algebra can be embedded into
some linear conformal algebra which is generated by a finite number of
currents and contains the considered $W$-algebra as a subalgebra in some
nonlinear basis. The currents of nonlinear algebra are
related by an {\it invertible} transformation to those of linear one and most
of the properties of the former and the theories constructed on its basis,
can be understood in a more simple way by studying its linear counterpart.
We called this linear algebra the linearizing algebra for the nonlinear one.

An idea to relate $W$-algebras and Lie algebras is also developed in \cite{TB},
however in our approach there is an essential difference: our linearizing
algebras are {\it conformal}, i.e. they contain Virasoro subalgebra and the
remainder of their currents are {\it primary} with respect to Virasoro stress
tensor. To underline this very important property of our linearizing
procedure we call it {\it conformal linearization}. Up to now the
explicit construction of conformal linearization has been carried out for
many examples of nonlinear (super)algebras \cite{KS,KSlin1,BKS,KSlin2,MR}
and in all these cases conformal linearizing algebras are more {\it efficient},
i.e. contain less currents than the algebras of Ref.~\cite{TB}. Besides being
a useful tool to construct a new more wide class of field
realizations of nonlinear algebras~\cite{KS,BKS}, these linear conformal
algebras provide a suitable framework for constructing new string
theories as well as studying the embeddings of the Virasoro string
in the $W$-type ones~\cite{BO,LPSX,LPX}.

In the present review we would like to demonstrate
that the conformal linearization is a general property inherent to
many nonlinear $W$-algebras from the $W(sl(N),H)$ series produced via the
Hamiltonian reduction constraints imposed on the affine $sl(N)$ currents,
associated with principal embedding of $sl(2)$ algebra into regular
subalgebra $H$ of $sl(N)$ \cite{BS,BTD,FRS}. We describe the heuristic method
of conformal linearization and present linearizing conformal algebras for
a wide class of $W$-(super)algebras.

Our approach \cite{KSlin1,KSlin2} is based on the {\it Conjecture} about the
relation between conformal linearizing algebra for $W(sl(N),H)$ and the
linearizing algebra for the algebra $\widetilde{W}$ obtained via special
Hamiltonian reduction applied to conformal linearizing algebra for
$W(sl(N),sl(2))$. The existence of correspondence between these linearizing
algebras seems reasonable if one remembers that $W(sl(N),sl(2))$ algebra
turns out to be
more universal than $W(sl(N),H)$ in the sense that the latter can be
generated via secondary Hamiltonian reductions from the former~\cite{DFRS,MR}.
The {\it Conjecture} as well as the method of construction of conformal
linearizing
algebras for $W(sl(N),sl(2))$ \cite{KSlin1,KSlin2} are inspired by the
analysis of simplest examples of conformal linearization for $W_3$
($W(sl(3),sl(3))$) and $W_3^{(2)}$ ($W(sl(3),sl(2))$) algebras \cite{KS}.
After finding the explicit form
of the conformal linearizing algebra for $W(sl(N),sl(2))$, we will show that
its different Hamiltonian reductions can be linearized using a slight
modification of the method used in \cite{TB} for the reductions of affine
algebra $sl(N)$. However due to the fact that we started from reductions
of conformal algebras, their linearizing algebras are also conformal.

We~ illustrate~ the general~ construction~ by~ the examples of
{}~$u(N)$-superconformal \cite{KB},
$W(sl(N),sl(2))$ \cite{BTD,Rom,F},
$W(sl(3),sl(3))$ \cite{W3}, $W(sl(4),sl(4))$ \cite{FL,Nahm}
as well as $W(sl(N),sl(3))$ algebras. The explicit formulas relating
linearizing and nonlinear conformal algebras for all these cases are given
and their new realizations are produced in this way.

An alternative approach to conformal linearization is developed
independently in \cite{MR} in the framework of the quantum secondary
Hamiltonian reduction. However the linearization of the \\ $W(sl(N),sl(2))$
algebras was not considered in \cite{MR}, because the method used there does
not allow fields with negative conformal weights which are necessary for
this purpose.

The review is organized as follows. In Section 2 we describe linearizing
conformal algebras for $W(sl(N),sl(2))$ as well as for their supersymmetric
counterpart - $u(N)$ superconformal algebras. In Section 3 we formulate
the main {\it Conjecture} and in the framework of BRST formalism we find the
general formulas for the currents of linearizing conformal algebras for
$W(sl(N),H)$. In Sections 4 and 5 we apply general approach of Section 3,
to $W(sl(N),sl(N))$ ($W_N$ \cite{FL}) and $W(sl(N),sl(3))$ series of
nonlinear algebras. And finally, in Section 6 we end with closing remarks.

\setcounter{equation}0
\section{Linearizing $W(sl(N+2),sl(2))$ and $u(N)$ superconformal
algebras.}

In this Section we construct conformal linearizing algebras for
$W(sl(N+2),sl(2))$ \cite{BTD,Rom,F} and $W(sl(N|2),sl(2))$
($u(N)$ superconformal \cite{KB}) (super)algebras \cite{KSlin1,KSlin2}.

Hereafter, some nonlinear redefinition
of the currents is called the change of the basis of the (non)linear algebra,
if (i) it is invertible and (ii) both it and its inverse are polynomial
in the currents and derivatives of the latter. A subset of the currents is
meant to form a (non)linear subalgebra of  given $W$-algebra if in some
basis this subset is closed; all the algebras related by (nonlinear)
transformations of the basis are treated as equivalent.

Let~ us~ start~ by~ reminding~ the~ operator~ product~
expansions~ (OPEs)~ for~ the
$W(sl(N+2),sl(2))$ and $u(N)$ superconformal algebras (SCAs).
The OPEs for these algebras can be written in a general uniform way
keeping in mind that the $W(sl(N+2),sl(2))$ algebra is none other than
$u(N)$ quasi-superconformal algebra (QSCA) \cite{BTD,Rom,F}\footnote{Strictly
speaking, the $W(sl(N+2),sl(2))$ algebra coincides with
$gl(N)$ QSCA. In what follows, we will not specify the real forms of algebras
and use the common term $u(N)$ QSCA.}.
Both $u(N)$ SCA and $u(N)$ QSCA have
the same number of generating currents: the stress tensor $T(z)$,
the $u(1)$ current $U(x)$, the $su(N)$ affine currents $J_{a}^{b}(x)$
$(1\leq a,b \leq N , \mbox{Tr}(J)=0)$ and two sets of
currents in the fundamental $G_a(x)$ and conjugated ${\bar G}^b(x)$
representations of $su(N)$. The currents $G_a(x),{\bar G}^b(x)$ are bosonic
for $u(N)$ QSCA and fermionic for $u(N)$ SCA. To distinguish between
these two cases we, following Ref. \cite{Rom}, introduce the parameter
$\epsilon$ equal to $1$ $(-1)$ for the QSCAs (SCAs) and write
the OPEs for these algebras in the following universal form:
\bea
T(z_1)T(z_2) & = & \frac{c/2}{z_{12}^4}+\frac{2T}{z_{12}^2}+
                   \frac{T'}{z_{12}} \quad , \quad
U(z_1)U(z_2)  =  \frac{c_1}{z_{12}^2} \; , \nonumber \\
T(z_1)J_{a}^{b}(z_2) & = & \frac{J_{a}^{b}}{z_{12}^2}+
                   \frac{{J_{a}^{b}}'}{z_{12}} \quad , \quad
T(z_1)U(z_2)  =  \frac{U}{z_{12}^2}+\frac{U'}{z_{12}} \;, \nonumber \\
T(z_1)G_a(z_2)& = & \frac{3/2 G_a}{z_{12}^2}+\frac{G_a}{z_{12}}\quad , \quad
T(z_1){\bar G}^a(z_2) =  \frac{3/2 {\bar G}^a}{z_{12}^2}+
                 \frac{{\bar G}^a}{z_{12}} \; , \nonumber \\
J_a^b(z_1)J_c^d(z_2) & = & (K-\epsilon -N)\frac{\delta_a^d\delta_c^b-
 \frac{1}{N}\delta_a^b\delta_c^d}{z_{12}^2}+
 \frac{\delta_c^b J_a^d-\delta_a^d J_c^b}{z_{12}} \; , \nonumber \\
U(z_1)G_a(z_2) & = & \frac{G_a}{z_{12}} \quad , \quad
U(z_1){\bar G}^a(z_2)  =  -\frac{{\bar G}^a}{z_{12}} \; , \nonumber \\
J_a^b(z_1)G_c(z_2) & = &
 \frac{\delta_c^b G_a -\frac{1}{N}\delta_a^b G_c}{z_{12}} \quad , \quad
J_a^b(z_1){\bar G}^c(z_2)  =  \frac{-\delta_a^c {\bar G}^b +
 \frac{1}{N}\delta_a^b {\bar G}^c}{z_{12}} \; \nonumber \\
G_a(z_1) {\bar G}^b(z_2) & = & \frac{2\delta_a^b c_2}{z_{12}^3}+
 \frac{2x_2\delta_a^b U + 2x_3 J_a^b}{z_{12}^2}+
 \frac{x_2\delta_a^b U' + x_3 {J_a^b}'+2x_5 (J_a^dJ_d^b)}{z_{12}}
          \nonumber \\
 & & + \frac{2x_4 (UJ_a^b)+\delta_a^b \left( x_1 (UU)- 2\epsilon T +
     2x_6 (J_d^eJ_e^d)\right) }{z_{12}} \; , \label{ope}
\eea
where the  central charges $c$ and parameters $x$ are defined by
\bea
c & = & \frac{-6\epsilon K^2 +(N^2+11\epsilon N +13)K-(\epsilon+N)
     (N^2+5\epsilon N+6)}{K} \; ,  \label{sasha2}
\eea
\bea
c_1 & = & \frac{N(2K-N-2\epsilon)}{2+\epsilon N} \quad , \quad
  c_2=\frac{(K-N-\epsilon )(2K-N-2\epsilon )}{K} \; ,  \nonumber \\
x_1 & = & \frac{(\epsilon +N)(2\epsilon+N)}{N^2K} \quad , \quad
 x_2 = \frac{(2\epsilon +N)(K-\epsilon-N)}{\epsilon NK} \quad , \quad
 x_3 = \frac{2K-N-2\epsilon }{K} \; , \nonumber \\
x_4 & = & \frac{2+\epsilon N}{NK} \quad , \quad
 x_5 = \frac{1}{K} \quad , \quad x_6 = \frac{1}{2\epsilon K} \; .
            \label{opecoeff}
\eea
The currents in the r.h.s. of OPEs \p{ope} are evaluated at the point $z_2$,
$z_{12}=z_1-z_2$ and the normal ordering in the nonlinear terms is
understood. Hereafter we will write only regular terms of OPEs.

The problem of construction of linear algebras for nonlinear ones can be
naturally divided in two steps. At the first step we need to find the
appropriate set of additional currents which linearize the given nonlinear
algebra. In other words, we should construct the linear algebra for extended
set of currents with the special relations between, for example, central
charges, conformal weights etc., so that it contains the nonlinear
algebra as a subalgebra in some nonlinear basis. At the second step, we
need to explicitly construct the transformation from the linear basis to a
nonlinear one. While the first step is highly non-trivial, the second one is
purely technical: one writes down the most general expressions in the
currents of linear algebra with arbitrary coefficients and conformal weights
appropriate to nonlinear algebra currents, and then fixes all the
coefficients from the OPEs of the given nonlinear algebra. In principle, if
we know from some consideration the true linearizing algebra we do not need
to know beforehand even OPEs of nonlinear algebra because all the information
about nonlinear algebra is encoded in the linear one. We could derive OPEs
of nonlinear algebra by demanding OPEs between the constructed general
expressions to form a closed set. Below in this Section we consider these two
steps for the algebras under consideration.

The main question we need to answer at first step in order to linearize the
$u(N)$ (Q)SCA \p{ope} is as to which minimal set of additional currents
must be added to $u(N)$ (Q)SCA to get extended linear conformal algebras
containing \p{ope} as subalgebras in some nonlinear basis. The idea of our
construction comes from the observation that the classical
$(K \rightarrow \infty )$ $u(N)$ (Q)SCA \p{ope} can be realized as left
shifts in the following coset space
\be
g = e^{\int \! dz {\bar Q}^a (z) G_a (z) } \;,\label{coset}
\ee
which is parametrized by $N$ parameters-currents ${\bar Q}^a(z)$ with
unusual conformal weights $-1/2$. In this case, the currents of
$u(N)$ (Q)SCA \p{ope} can be realized in terms of ${\bar Q}^a(z)$ and
their conjugated momenta $G_a(z) = \delta /\delta {\bar Q}^a$
with OPEs
\be
G_a(z_1) {\bar Q}^b(z_2)  =  \frac{\delta_a^b}{z_{12}} \;, \label{gam}
\ee
as well as the currents of the maximal {\it linear} subalgebra ${\cal H}_N$
\be
{\cal H}_N  = \left\{ T , U, J_a^b , {\bar G}^a \right\} \; . \label{glin}
\ee
Moreover, a realization of a given current from $u(N)$ (Q)SCA will contain
some linear term belonging to the set of currents
$\left\{ T , U, J_a^b , {\bar G}^a, {\bar Q}^a(z), G_a(z) \right\}$  which
form the linear algebra. So actually such a realization describes the change
of the basis from this linear algebra to the nonlinear algebra $u(N)$ (Q)SCA
extended by the currents ${\bar Q}^a(z)$ .

Though the situation in quantum case is more difficult, it still seems
reasonable to try to generalize this classical picture to the quantum
case, i.e. to extend the $u(N)$ (Q)SCA  \p{ope} by $N$
additional currents ${\bar Q}^a(z)$ with conformal weights
$-1/2$, especially keeping in mind that the current with just this conformal
weight appears in the linearization of $W_3^{(2)}$ algebra \cite{KS}.

Fortunately, ~this~ extension~ is sufficient~ to construct~ the
linearizing algebra~ for the $u(N)$~(Q)SCA in the quantum case also.
Without going into details, let us write down the set of OPEs for
this linear algebra with the currents
$\left\{ T(z),U(z),J_a^b(z),G_a(z), \widetilde{\overline G}{}^a(z),
{\bar Q}^a(z)\right\}$, which we denote as $(Q)SCA_N^{lin}$ algebra
\bea
T(z_1)T(z_2) & = & \frac{c/2}{z_{12}^4}+\frac{2T}{z_{12}^2}+
                   \frac{T'}{z_{12}} \quad , \quad
U(z_1)U(z_2)  =  \frac{c_1}{z_{12}^2} \; , \nonumber \\
T(z_1)J_{a}^{b}(z_2) & = & \frac{J_{a}^{b}}{z_{12}^2}+
                   \frac{{J_{a}^{b}}'}{z_{12}} \quad , \quad
T(z_1)U(z_2)  =  \frac{U}{z_{12}^2}+\frac{U'}{z_{12}} \;, \nonumber \\
T(z_1)G_a(z_2)& = & \frac{3/2 G_a}{z_{12}^2}+\frac{G_a}{z_{12}}\quad , \quad
T(z_1)\widetilde{\overline G}{}^a(z_2) =
     \frac{3/2 \widetilde{\overline G}{}^a}{z_{12}^2}+
                 \frac{\widetilde{\overline G}{}^a}{z_{12}}\; , \nonumber \\
T(z_1){\bar Q}^a(z_2) & = & \frac{-1/2 {\bar Q}^a}{z_{12}^2}+
                 \frac{{\bar Q}^a}{z_{12}}\; , \nonumber \\
J_a^b(z_1)J_c^d(z_2) & = & (K-\epsilon -N)\frac{\delta_a^d\delta_c^b-
 \frac{1}{N}\delta_a^b\delta_c^d}{z_{12}^2}+
 \frac{\delta_c^b J_a^d-\delta_a^d J_c^b}{z_{12}} \; , \nonumber \\
U(z_1)G_a(z_2) & = & \frac{G_a}{z_{12}} \quad , \quad
U(z_1)\widetilde{\overline G}{}^a(z_2)  =
     -\frac{\widetilde{\overline G}{}^a}{z_{12}}  \quad , \quad
U(z_1){\bar Q}^a(z_2)  =  -\frac{{\bar Q}^a}{z_{12}} \; , \nonumber \\
J_a^b(z_1)G_c(z_2) & = &
 \frac{\delta_c^b G_a -\frac{1}{N}\delta_a^b G_c}{z_{12}} \quad , \quad
J_a^b(z_1)\widetilde{\overline G}{}^c(z_2)  =
       \frac{-\delta_a^c \widetilde{\overline G}{}^b +
 \frac{1}{N}\delta_a^b \widetilde{\overline G}{}^c}{z_{12}} \; , \nonumber \\
J_a^b(z_1){\bar Q}^c(z_2) & = & \frac{-\delta_a^c {\bar Q}^b +
 \frac{1}{N}\delta_a^b {\bar Q}^c}{z_{12}} \; , \nonumber \\
G_a(z_1) {\bar Q}^b(z_2) & = & \frac{\delta_a^b}{z_{12}}
 \quad , \quad G_a(z_1) \widetilde{\overline G}{}^b(z_2)  = \mbox{regular}
 \;. \label{linal1}
\eea
Here the central charges $c$ and $c_1$ are the same as in \p{opecoeff} and
the currents $G_a(z),\widetilde{\overline G}{}^a(z)$ and ${\bar Q}^a(z)$ are
bosonic (fermionic) for $\epsilon = 1$ $(-1)$.

At the second step, in order to prove that the linear algebra
$(Q)SCA_N^{lin}$ \p{linal1} contains  $u(N)$~(Q)SCA  \p{ope} as a subalgebra,
let us perform the following {\it invertible} nonlinear transformation
to the new basis
$\left\{ T(z),U(z),J_a^b(z),G_a(z),{\bar G}^a(z),{\bar Q}^a(z)\right\}$,
where the "new" current ${\bar G}^a(z)$ is defined as
\bea
{\bar G}^a & = & \widetilde{\overline G}{}^a + y_1 {\bar Q}^a{}''+
 y_2 (J_b^a{\bar Q}^b{}') + y_3 (U{\bar Q}^a{}')+y_4 ({J_b^a}'{\bar Q}^b)+
       y_5 (U'{\bar Q}^a)+y_6 (T{\bar Q}^a)\nonumber \\
 & & + y_7(J_b^cJ_c^a{\bar Q}^b)+
 y_8(J_b^cJ_c^b{\bar Q}^a)+ y_9 (UJ_b^a{\bar Q}^b)+ y_{10} (UU{\bar Q}^a) +
      y_{11}(J_b^cG_c{\bar Q}^b{\bar Q}^a)  \nonumber \\
 & & + y_{12}(J_b^aG_c{\bar Q}^c{\bar Q}^b)+y_{13}(G_b'{\bar Q}^b{\bar Q}^a)+
    y_{14}(G_b{\bar Q}^b{}'{\bar Q}^a)+
    y_{15}(G_b{\bar Q}^b{\bar Q}^a{}') \nonumber \\
 & & + y_{16}(G_bG_c{\bar Q}^b{\bar Q}^c{\bar Q}^a) +
    y_{17}(UG_b{\bar Q}^b{\bar Q}^a) \; , \label{tr1}
\eea
and the coefficients $y_1-y_{17}$ are defined as
\bea
y_1 & = & 2K \quad , \quad y_2 = 4 \quad , \quad
  y_3=\frac{2(2+\epsilon N)}{N} \quad , \quad
  y_4=\frac{2(K-\epsilon -N)}{K} \;, \nonumber \\
y_5 & = & \frac{(K-\epsilon-N)(2+\epsilon N)}{NK} \quad , \quad
  y_6=-2\epsilon \quad , \quad y_7=\frac{2}{K} \quad , \quad
  y_8=\frac{2}{\epsilon K} \quad , \quad y_9=\frac{2(2+\epsilon N)}{NK}
        \; , \nonumber \\
y_{10} & = & \frac{(\epsilon+N)(2\epsilon+N)}{N^2K} \quad , \quad
 y_{11} = y_{12}=\frac{2}{K} \quad , \quad
 y_{13}= \frac{2(K-N-2\epsilon )}{K}\quad , \quad y_{14}=4 \; \nonumber \\
y_{15} & = & 2 \quad , \quad y_{16}=\frac{2}{\epsilon K} \quad , \quad
 y_{17}=\frac{2(2+\epsilon N)}{NK} \; . \label{sing}
\eea
Now it is a matter of straightforward (though tedious) calculation to check
that OPEs for the subset of currents
$\left\{ T(z),U(z),J_a^b(z),G_a(z)\right\}$ and ${\bar G}^a(z)$ \p{tr1}
coincide with the basic OPEs of the $u(N)$ (Q)SCA \p{ope}.

Thus, we have shown that the linear algebra $(Q)SCA_N^{lin}$ \p{linal1}
contains  $u(N)$ (Q)SCA as a subalgebra in the nonlinear basis.

We close this Section with a few comments.

First of all, the pairs of currents
 $G_a(z)$ and ${\bar Q}^a(z)$ (with conformal weights
equal to $3/2$ and $-1/2$, respectively) in \p{linal1} look like
``ghost--anti-ghost'' fields and so $(Q)SCA_N^{lin}$  algebra \p{linal1}
can be simplified by means of the standard ghost decoupling
transformations
\bea
U & = & {\widetilde U}-\epsilon (G_a{\bar Q}^a) \; , \nonumber \\
J_a^b & = & {\widetilde J}{}_a^b - \epsilon (G_a{\bar Q}^b) +
                \delta_a^b\frac{\epsilon}{N}(G_c{\bar Q}^c) \; , \nonumber \\
T & = & {\widetilde T} +\frac{1}{2}\epsilon (G_a'{\bar Q}^a)
       +\frac{3}{2}\epsilon (G_a{\bar Q}^a{}')
        -\frac{\epsilon (2+ \epsilon N)}{2K} {\widetilde U}{}' \; ,
                                              \label{ghosts}
\eea
where the term with the derivative of the current ${\widetilde U}$ is added
to ensure primarity of ${\widetilde U}$ with respect to the new stress
tensor ${\widetilde T}$.
In this new basis the algebra $(Q)SCA_N^{lin}$ splits into the direct sum
\be
(Q)SCA_N^{lin}=\Gamma_N \oplus \widetilde{(Q)SCA}{}_N^{lin}  \label{best}
\ee
of the ghost--anti-ghost algebra $\Gamma_N=\left\{ {\bar Q}^a, G_b \right\}$
with the OPEs \p{gam} and the algebra
$\widetilde{(Q)SCA}{}_N^{lin}=\left\{ {\widetilde T},{\widetilde U},
{\widetilde J}{}_a^b,\widetilde{\overline G}{}^a\right\}$ with the following
set of OPEs
\bea
{\widetilde T}(z_1){\widetilde T}(z_2) & = &
 \frac{-6\epsilon K^2 +(N^2+13)K -(N^3-N+6\epsilon )}{2K\; z_{12}^4}+
  \frac{2{\widetilde T}}{z_{12}^2}+
                   \frac{{\widetilde T}'}{z_{12}} \quad , \nonumber \\
{\widetilde U}(z_1){\widetilde U}(z_2) & = &
  \left(\frac{2NK}{2+\epsilon N}\right)\frac{1}{z_{12}^2} \; , \;
{\widetilde T}(z_1){\widetilde J}{}_{a}^{b}(z_2) =
 \frac{{\widetilde J}{}_{a}^{b}}{z_{12}^2}+
                   \frac{{\widetilde J}{}_{a}^{b}{}'}{z_{12}} \; ,
                      \nonumber \\
{\widetilde T}(z_1){\widetilde U}(z_2)  & = &
    \frac{{\widetilde U}}{z_{12}^2}+\frac{{\widetilde U}'}{z_{12}} \;,
               \nonumber \\
{\widetilde T}(z_1)\widetilde{\overline G}{}^a(z_2) & = &
 \left( \frac{3}{2}+\frac{\epsilon (2+\epsilon N)}{2K}\right)
            \frac{\widetilde{\overline G}{}^a}{z_{12}^2}+
                 \frac{\widetilde{\overline G}{}^a}{z_{12}}\; , \nonumber \\
{\widetilde J}{}_a^b(z_1){\widetilde J}{}_c^d(z_2) & = &
  (K-N)\frac{\delta_a^d\delta_c^b-\frac{1}{N}\delta_a^b\delta_c^d}{z_{12}^2}+
 \frac{\delta_c^b {\widetilde J}{}_a^d-
          \delta_a^d {\widetilde J}{}_c^b}{z_{12}} \; , \nonumber \\
{\widetilde U}(z_1)\widetilde{\overline G}{}^a(z_2)  & =  &
     -\frac{\widetilde{\overline G}{}^a}{z_{12}}  \; , \;
{\widetilde J}{}_a^b(z_1)\widetilde{\overline G}{}^c(z_2)  =
       \frac{-\delta_a^c \widetilde{\overline G}{}^b +
 \frac{1}{N}\delta_a^b \widetilde{\overline G}{}^c}{z_{12}} \; , \nonumber \\
\widetilde{\overline G}{}^a(z_1) \widetilde{\overline G}{}^b(z_2)  & = &
  \mbox{regular} \;, \label{linal2}
\eea

Secondly, note that the linear algebra
$\widetilde{(Q)SCA}{}_N^{lin}$ \p{linal2}
has the same number of currents and the same structure relations as the
maximal linear subalgebra ${\cal H}_N$ \p{glin} of $u(N)$ (Q)SCA \p{ope},
but with the "shifted" central charges and
conformal weights. It is of importance that the central charges and
conformal weights are strictly related as in \p{linal2}\footnote{Let
us point out that Jacobi identities for the set of currents
$\left\{ {\widetilde T},{\widetilde U},
{\widetilde J}{}_a^b,\widetilde{\overline G}{}^a\right\}$
do not fix neither central charges nor the conformal weight of
$\widetilde{\overline G}{}^a$.}.
Otherwise, with  another relation between these parameters, we would
never find the $u(N)$ (Q)SCA \p{ope} in
$(Q)SCA{}_N^{lin}$.
Thus, our starting assumption about the structure of linear algebra for
$u(N)$ (Q)SCA  coming from the classical coset realization approach, proved
to be correct, modulo shifts of central charges and conformal weights.

Thirdly, let us remark that among the $u(N)$ (Q)SCAs there are many
(super)algebras which are well known under other names.
For example\footnote{To avoid the singularity
in \p{opecoeff} at $\epsilon=-1,N=2$ one should firstly rescale the current
$U\rightarrow \frac{1}{\sqrt{2+\epsilon N}}U$ and then put $\epsilon=-1,N=2$
\cite{KB}.}:
\bea
(Q)SCA ( \epsilon=1,N=1) & \equiv & W_3^{(2)}  \quad
\mbox{Ref.\cite{P,B}}\; , \nonumber \\
(Q)SCA ( \epsilon=-1,N=1) & \equiv & N=2\; SCA \quad \
\mbox{Ref.\cite{A}} \; , \nonumber \\
(Q)SCA ( \epsilon=-1,N=2) & \equiv & N=4
\; SU(2)\; SCA  \quad
\mbox{Ref.\cite{A}} \; . \nonumber
\eea

Finally, the linear algebra $\widetilde{(Q)SCA}{}_N^{lin}$ \p{linal2}
is homogeneous in the currents ${\widetilde{\overline G}{}^a}$, so they
are null fields. Evidently we could consistently put them equal to zero,
${\widetilde{\overline G}{}^a} = 0$, and be left with the Miura realization
\p{tr1} of the  $u(N)$ (Q)SCA \p{ope} in terms of currents
${\widetilde T_{Vir}}$, ${\widetilde U}$, ${\widetilde J}{}_a^b$,
${\bar Q}^a$ and $G_b$, where we introduced decoupling basis in
$\widetilde{(Q)SCA}{}_N^{lin}$ algebra with the new stress tensor
${\widetilde T_{Vir}}$
\bea
{\widetilde T_{Vir}} & = & {\widetilde T} -
       \frac{1}{2K} {\widetilde J}{}_a^b{\widetilde J}{}_b^a -
       \frac{2+\epsilon N}{4NK}{\widetilde U}{\widetilde U} \; ,
                                              \label{sasha1}
\eea
commuting with all other currents and having the following central charge
$c_{Vir}$
\bea
c_{Vir} =1-6\frac{(K-1)^2}{K} \; .  \label{vir}
\eea
In this basis at ${\widetilde{\overline G}{}^a} = 0$
the $\widetilde{(Q)SCA}{}_N^{lin}$ algebra \p{linal2} splits
in a direct sum of Virasoro, $u(1)$ and $sl(N)$ affine algebras.
The values of $c_{Vir}$ corresponding to the minimal models of Virasoro
algebra~\cite{min} at
\be
K=\frac{p}{q} \Rightarrow c_{Vir}=1-6\frac{(p-q)^2}{pq}  \label{level}
\ee
induce the following spectrum for central charge $c$ \p{sasha2}
of $u(N)$ (Q)SCA \p{ope}
\bea
c & = & \frac{-6\epsilon p^2 +(N^2+11\epsilon N +13)pq-(\epsilon+N)
     (N^2+5\epsilon N+6)q^2}{pq} \; .  \label{sasha3}
\eea
One can check for the particular cases of $N=2$ superconformal and
$W_3^{(2)}$ algebras that corresponding spectrum contains the
spectrum of minimal models for these algebras \cite{BFK,B}.
Moreover, as we will show in the Section 4 for the case of $W_N$ algebra,
spectrum of central charge for its minimal models can be reproduced by
minimal models of this Virasoro algebra. So it seems reasonable to suppose
that this property will remain also for the whole series of $u(N)$ (Q)SCAs.
Nevertheless, our conjecture must be checked by the standard methods.

Besides these simplest realizations there are ones with null currents which
are realized in terms of free fields by a non-vanishing operator.
In Section 4.1 we will discuss such realizations for $W_3$ algebra.

In the simplest case of $W_3^{(2)}$ algebra, the linear
$\widetilde{QSCA}{}_1^{lin}$ algebra \p{linal2} coincides with the
linear algebra $W_3^{lin}$ \cite{KS} for $W_3$. For general $N$ the
situation is more complicated. This question will be discussed in the next
Section.

\section{Secondary linearization of $W(sl(N+2),H)$ algebras.}

In this Section following Refs. \cite{KSlin1,KSlin2} we demonstrate that
the linear algebra
$QSCA_N^{lin}$ \p{linal1} constructed in the previous Section gives the
hints how to find the linearizing algebras for many other $W$-type algebras
which can be obtained from the $W(sl(N+2),sl(2))$ ($u(N)$ QSCAs) via the
secondary Hamiltonian reduction \cite{DFRS}.

The $W(sl(N+2),sl(2))$ algebra,
which have been linearized in the previous Section, can be obtained through
the primary Hamiltonian reduction with {\it minimal} set of constraints from
the affine $sl(N+2)$ algebras \cite{BTD,Rom,F} and so at a fixed value of $N$
it contains the {\it maximally} possible set of the currents.
The full set of constraints on the currents of $sl(N+2)$ algebra which
yield $W(sl(N+2),sl(2))$ read
\be
\left(
  \begin{array}{cc|cccc}
  U &               T & {\overline G}{}^1 & {\overline G}{}^2 & \ldots &
                                     {\overline G}{}^N \\
  1 &               0 & 0   & 0   & \ldots & 0 \\ \hline
  0 & G_1 &     &     &        &   \\
  0 & G_2 &     &     &        &   \\
  \vdots & \vdots &  \multicolumn{4}{c}{ sl(N) - \frac{\delta_a^b}{N}U}   \\
  0 & G_N &     &     &        &
  \end{array}
\right) . \label{ss1}
\ee
The $W(sl(N+2),sl(2))$ algebras, forming in themselves a particular class
of $W$-algebras with quadratic nonlinearity, are at the same time universal
in the sense that a lot of other
$W$-algebras with higher nonlinearity can be obtained from them via the
secondary Hamiltonian reduction (e.g., $W_N$ algebras, etc.) \cite{DFRS}.

Let us consider a set of possible secondary Hamiltonian reductions of
$W(sl(N+2),sl(2))$ algebra \p{ss1}, \p{ope} to nonlinear algebras
which at the same time could be also produced via primary Hamiltonian
reductions of affine $sl(N+2)$ algebra and belong to
the $W(sl(N+2),H)$ series\footnote{Let us remind that by definition the
$W(sl(N),H)$ algebra is a nonlinear algebra produced via the primary
Hamiltonian reduction constraints imposed on the affine $sl(N)$ currents,
associated with principal embedding of $sl(2)$ algebra into regular
subalgebra $H$ of $sl(N)$, and this series forms the complete set of
nonlinear algebras associated with different $sl(2)$ embeddings into
$sl(N)$ \cite{BS,BTD,FRS}.}. These are introduced by imposing the following
constraints on the $W(sl(N+2),sl(2))$ currents
\bea
G_1 = 1 \quad , \quad
      G_2=\ldots =G_N=0 \quad ,& & \label{ss2}\\
 \left. sl(N)\right|_{sl(2)} \quad , & & \label{ss3}
\eea
where we denoted as $\left. sl(N)\right|_{sl(2)}$ the set of constraints
on the $sl(N)$ currents associated with an arbitrary embedding of $sl(2)$
algebra into $sl(N)$ subalgebra of $W(sl(N+2),sl(2))$.

The main {\it Conjecture} we will keep to in this Section is as follows
\begin{quote}\it
To~ find~ the linearizing~ algebra~ for~ a given~ nonlinear~ $W$~ algebra~
related to
$W(sl(N+2),sl(2))$ through the Hamiltonian reduction constraints \p{ss2}
and/or \p{ss3}, one should apply the same reduction to its linearizing
algebra ${QSCA}{}_N^{lin}$ \p{best} and then linearize the resulting
algebra.
\end{quote}

Taking into account that ${QSCA}{}_N^{lin}$ has the structure of direct sum
\p{best} of the ghost--anti-ghost $\Gamma_N=\left\{ {\bar Q}^a, G_b \right\}$
\p{gam} and $\widetilde{(Q)SCA}{}_N^{lin}=\left\{ {\widetilde T},
{\widetilde U},{\widetilde J}{}_a^b,\widetilde{\overline G}{}^a\right\}$
\p{linal2} algebras, as well as that the currents ${\bar Q}^a$ are the gauge
degree of freedom for the gauge transformations produced by constraints
\p{ss2}, one can get
another, but equivalent following form of {\it Conjecture}

\begin{quote}\it
i) To~ find~ the~ linearizing~ algebra~ for~ a given~ nonlinear~
$W$~algebra~ related to
$W(sl(N+2),sl(2))$ through the Hamiltonian reduction constraints
\p{ss2} and \p{ss3}, one should apply the reduction \p{ss3} to the linear
algebra $\widetilde{QSCA}{}_N^{lin}$ \p{linal2} and then linearize the
resulting algebra. \\
ii) The algebra $\widetilde{QSCA}{}_N^{lin}$ itself is the linearizing
algebra for the reduction \p{ss2}, i.e. for $W(sl(N+2),sl(3))$ algebra. \\
iii) Linearizing algebra for reduction \p{ss3} has the structure of direct
sum of the algebra $\Gamma_N$ and linearizing algebra for reductions \p{ss3}
of algebra $\widetilde{QSCA}{}_N^{lin}$ \p{linal2}.
\end{quote}

Thus, in fact {\it Conjecture} reduces the
problem of conformal linearization of the algebra $W$ obtained from the
nonlinear algebra $W(sl(N+2),sl(2))$ through the full set of the Hamiltonian
reduction constraints \p{ss2} and/or \p{ss3} (i.e. $W$ belongs to
$W(sl(N+2),H)$ series) to the problem of linearization of the algebra
$\widetilde{W}$ obtained from the more simple {\it linear} algebra
$\widetilde{QSCA}{}_N^{lin}$ by imposing the relaxed set \p{ss3}. At present,
we are not aware of the rigorous proof of this assumption, but it works in
many examples: $W_3$, $W_4$ algebras (see Subsections 4.1 and 4.2) and in
Section 5 we will prove the point ii) of {\it Conjecture} concerning
$W(sl(N+2),sl(3))$ algebras.

Of course, the secondary Hamiltonian reduction \p{ss3}, being applied to
$\widetilde{QSCA}{}_N^{lin}$, gives rise to a nonlinear algebra. However,
the problem of its linearization as we will show below can be reduced to
the linearization of reduction \p{ss3} applied to the affine subalgebra
$sl(N)\subset \widetilde{QSCA}{}_N^{lin}$, which was already constructed
in \cite{TB}. The resulting algebra will be just linearizing algebra for
the nonlinear algebra we started with.

Let us briefly discuss the explicit construction of the
linearizing algebra $W^{lin}$ for the nonlinear
algebra $\widetilde{W}$ obtained from $\widetilde{QSCA}{}_N^{lin}$
via the Hamiltonian reduction constraints \p{ss3}.

Let ${\cal J}$ be a current corresponding to the Cartan element $t_0$ of
$sl(2)$ subalgebra. With respect to the adjoint action of $t_0$
the $sl(N)$ algebra can be decomposed into eigenspaces of  $t_0$
with positive, null and negative  eigenvalues $h_a$
\be
sl(N) = \left( sl(N) \right)_{-} \oplus \left( sl(N) \right)_{0} \oplus
         \left( sl(N) \right)_{+} \equiv
        \begin{array}[t]{c}
            \oplus \\
             h_a \end{array}
            \left( sl(N) \right)_{h_a}
\quad . \label{f2}
\ee
(In this Section below, the latin indices $(a,b)$ run over the whole
$sl(N)$, Greek indices $(\alpha,\beta )$ run over
$\left( sl(N) \right)_{-}$ and the barred Greek ones $(\bar\alpha,\bar\beta )$
over $\left( sl(N) \right)_{0} \oplus \left( sl(N) \right)_{+} $ .)
The Hamiltonian reduction associated with the decomposition \p{f2} can be
performed by putting the appropriate constraints
\be
J_{\alpha}-\chi_{\alpha}=0 \quad , \quad
  \chi_{\alpha}\equiv \chi (J_{\alpha}) \label{sor1}
\ee
on the currents $J_{\alpha}$ from $\left( sl(N) \right)_{-}$ \cite{BS,BTD}.
These constraints are first class for
integral gradings\footnote{Let us remind, that the half-integer gradings
can be replaced by integer ones, leading to the same reduction \cite{BS}.},
which means that BRST formalism can be used.

In order to impose the constraints \p{sor1} in the framework of BRST
approach one can introduce the fermionic ghost--anti-ghost pairs
$( b_{\alpha},c^{\alpha} )$ with ghost numbers $(-1, 1)$, respectively,
for each current with the negative eigenvalues $h_{\alpha}$, and with
standard OPEs
\be
  c^{\alpha}(z_1)b_{\beta}(z_2) = \frac{\delta^{\alpha}_{\beta}}{z_{12}}
 \quad ,  \label{fermgh}
\ee
and the BRST charge
\be
Q_{BRST} = \int dz J_{BRST}(z) = \int dz
  \left( (J_{\alpha}-\chi (J_{\alpha}))c^{\alpha} -\frac{1}{2}
   f_{\alpha,\beta }^{\gamma}b_{\gamma}c^{\alpha}c^{\beta}\right) \;,
 \label{brst}
\ee
which coincides with that given in the paper \cite{TB}.
The currents $\left\{ {\widetilde T},{\widetilde U},
{\widetilde J}{}_a^b,\widetilde{\overline G}{}^a\right\}$
of the algebra $\widetilde{QSCA}_N^{lin}$ and the ghost fields
$\left\{b_{\alpha},c^{\alpha}\right\}$ form the BRST complex,
graded by the ghost number.
The $\widetilde{W}$ algebra is defined in this approach as the algebra of
operators generating the BRST charge null cohomology of this complex.

Following \cite{TB}, let us introduce the "hatted" currents
${\widehat J}{}_a$ :
\be
{\widehat J}{}_a  =  {\widetilde J}{}_a+
            \sum_{\beta,\gamma}
        f_{a,\beta}^{\gamma}b_{\gamma}c^{\beta} \; ,
                        \label{hat1}
\ee
where $f_{a,\beta}^{\gamma}$
are structure constants of $sl(N)$ in the basis \p{f2}.
As shown in \cite{TB}, the $W$-algebras, associated with
the reductions \p{ss3} of the affine $sl(N)$ can be embedded into linear
algebras formed by the subset of non-constrained currents
${\widehat J}{}_{\overline \alpha}$.

In contrast to the sl(N) algebra, our algebra $\widetilde{QSCA}{}_N^{lin}$
contains, besides the $sl(N)$ currents ${\widetilde J}{}_a^b$, three
additional ones ${\widetilde T},{\widetilde U},{\widetilde{\overline G}}{}^a$.
Fortunately, the presence of these currents create no new problems
while we construct a linearizing algebra for the reduction of
$\widetilde{QSCA}{}_N^{lin}$ with the BRST charge \p{brst}.
Namely, the improved stress-tensor ${\widehat T}$ with respect to which
$J_{BRST}$ in  eq. \p{brst} is a spin 1 primary current can be easily
constructed
\be
{\widehat T}  =  {\widetilde T} +{\cal J}' +
     \sum_{\alpha} \left\{ -(1+h_{\alpha})
         b_{\alpha}c^{\alpha}{}' - h_{\alpha}b_{\alpha}'c^{\alpha}
                \right\}  \; , \label{TT}
\ee
so together with the zero ghost number current ${\widetilde U}$ it
commutes with $Q_{BRST}$ and belongs to both $\widetilde{W}$ and
its linearizing algebra we are searching for.
As regards the current ${\widetilde{\overline G}}{}^i$, one could check
that it extends the complex generated
by the currents ${\widehat J}_a,b_{\alpha},c^{\beta}$ with  preserving
the structure of the BRST subcomplexes of the paper \cite{TB},
and forms, together with non-constrained currents
${\widehat J}_{\overline \alpha}$ and $c^{\alpha}$, a reduced BRST
subcomplex and subalgebra which does not contain the currents with
negative ghost numbers. Hence, following the same arguments which are given
in \cite{TB}\footnote{We don't reproduce here all details and refer a reader
to the original paper \cite{TB}.}, we can conclude that besides the
non-constrained currents ${\widehat J}_{\overline \alpha}$, the currents
${\widetilde{\overline G}}{}^i$ also belong to the set of linearizing algebra
currents for the $\widetilde{W}$ algebra and the last one closes not only
modulo BRST exact operators, but in its own right.

Thus, the subset of the non-constrained currents
${\widehat J}_{\overline \alpha}$ \p{hat1}, stress tensor ${\widehat T}$
\p{TT} and the currents
\be
{\widehat U} \equiv {\widetilde U} \quad ,\quad
 {\widehat{\overline G}} \equiv {\widetilde{\overline G}}{}^i \label{sor2}
\ee
form the conformal linearizing algebra $W^{lin}$ for the nonlinear algebra
$\widetilde{W}$. According to our main {\it Conjecture}, it forms, modulo
algebra $\Gamma_N$, the linearizing algebra for $W$ algebra obtained from
$W(sl(N+2),sl(2))$ through the secondary Hamiltonian reduction associated
with constraints \p{ss2} and/or \p{ss3}.

Let us close this Section with a few remarks.

All results of this Section can be naturally generalized to the case of
$W(sl(N|2),H)$ superalgebras. Thus all the considerations given here will
be valid, if altogether $W(sl(N+2),H)$ algebras are replaced by
$W(sl(N|2),H)$ superalgebras and in the sets of constraints \p{ss2} and
\p{ss3} only \p{ss3} are kept (constraints \p{ss2} are forbidden for
superalgebras due to fermionic statistic of current $G_1$), so in this case
only the point iii) of {\it Conjecture} survives.

Conformal linearizing algebras constructed in this Section are more efficient
compared to non-conformal linearizing algebras constructed in \cite{TB} for
corresponding primary Hamiltonian reductions of $sl(N+2)$ affine algebra.
Simple comparison shows that a given conformal linearizing algebra contains
by $N+1$ (by $1$) currents less than the one from \cite{TB} for the algebras
corresponding to points i), ii) (to point  iii) ) of {\it Conjecture}.

In the next Sections we will illustrate the general construction of this
Section by the examples of $W(sl(N),sl(N))$ ($W_N$) and $W(sl(N),sl(3))$
algebras.

\setcounter{equation}0
\section{\bf Linearizing $W_N$ algebras.}

In this Section we apply the general procedure described in the previous
Section 3 to the case of the principal embedding of $sl(2)$ into $sl(N)$
algebra to construct the linearizing algebras $W_N^{lin}$
\cite{KSlin1,KSlin2} for $W_N$ ($W(sl(N),sl(N))$) \cite{FL} algebras
corresponding to the point i) of {\it Conjecture}.

For the principal embedding of $sl(2)$ into $sl(N)$ with the currents
${\widetilde J}{}_a^b, (1\leq a,b \leq N, Tr(J) = 0)$, current ${\cal J}$
which correspond to the Cartan element of $sl(2)$ is defined to be

\be
  {\cal J}= - \sum_{m=1}^{N-1} m {\widetilde J}{}_{N-m}^{N-m} \quad ,
\label{cartan}
\ee
and the decomposition of affine algebra $sl(N)$ reads as follows
\bea
\left( sl(N) \right)_{-}  \propto  \left\{{\widetilde J}{}_a^b ,
( 2\leq b \leq N , 1\leq a <b ) \right\}, & & \nonumber \\
\left( sl(N) \right)_{0} \oplus \left( sl(N) \right)_{+} \propto
         \left\{{\widetilde J}{}_a^b , ( 1\leq a \leq N-1 ,
          a\leq b \leq N ) \right\}  \quad , \label{deco}
\eea
i.e. $\left( sl(N) \right)_{-}$ consists of those entries of the $N\times N$
current matrix which stand below the main
diagonal, and the remainder  just constitutes the subalgebra
$\left( sl(N) \right)_{0} \oplus \left( sl(N) \right)_{+}$.

Now, using general expressions of previous Section for the linearizing
algebra currents \p{hat1}~--~\p{sor2} with principal gradation \p{cartan},
\p{deco} and OPEs \p{linal2}, \p{fermgh}, and introducing the new
stress tensor ${\widehat {\cal T}}$
\bea
{\widehat {\cal T}} & = & {\widehat T} + \frac{(N+2)(K-1)}{2K}{\widehat U}' \;,
\eea
we are able to explicitly write OPEs for the currents of conformal linearizing
algebra $W_{N+2}^{lin}$ which contains the $W_{N+2}$ algebra as a subalgebra:
\bea
{\widehat {\cal T}}(z_1){\widehat {\cal T}}(z_2) & = &
 \frac{(N+1)\left( 1-(N+2)(N+3)\frac{(K-1)^2}{K}\right)}{2z_{12}^4}+
  \frac{2{\widehat {\cal T}}}{z_{12}^2}+
                   \frac{{\widehat {\cal T}}'}{z_{12}} \quad , \nonumber \\
{\widehat U}(z_1){\widehat U}(z_2) & = &
  \left(\frac{2NK}{2+N}\right)\frac{1}{z_{12}^2} \; , \nonumber \\
{\widehat {\cal T}}(z_1){\widehat J}_a^b(z_2) & = &
       \frac{(N+1-2a)(K-1)\delta_a^b}{z_{12}^3}+
        \frac{(b-a+1){\widehat J}_a^b}{z_{12}^2}+
                   \frac{{\widehat J}_a^b{}'}{z_{12}} \; ,
                      \nonumber \\
{\widehat {\cal T}}(z_1){\widehat U}(z_2)  & = & -\frac{2N(K-1)}{z_{12}^3}+
    \frac{{\widehat U}}{z_{12}^2}+\frac{{\widehat U}'}{z_{12}} \;,
               \nonumber \\
{\widehat {\cal T}}(z_1)\widehat{\overline G}{}^i(z_2) & = &
            \frac{(i+2)\widehat{\overline G}{}^i}{z_{12}^2}+
                 \frac{\widehat{\overline G}{}^i{}'}{z_{12}}\; , \nonumber \\
{\widehat J}_a^b(z_1){\widehat J}_c^d(z_2) & = &
  K\frac{\delta_a^d\delta_c^b-
       \frac{1}{N}\delta_a^b\delta_c^d}{z_{12}^2}+
        \frac{\delta_c^b {\widehat J}_a^d-
          \delta_a^d {\widehat J}_c^b}{z_{12}} \; , \nonumber \\
{\widehat U}(z_1)\widehat{\overline G}{}^i(z_2)  & =  &
     -\frac{\widehat{\overline G}{}^i}{z_{12}}  \; , \;
{\widehat J}_a^b(z_1)\widehat{\overline G}{}^i(z_2)  =
       \frac{-\delta_a^i \widehat{\overline G}{}^b +
 \frac{1}{N}\delta_a^b \widehat{\overline G}{}^i}{z_{12}} \; , \nonumber \\
\widehat{\overline G}{}^i(z_1) \widehat{\overline G}{}^j(z_2)  & = &
  \mbox{regular} \;, \label{linal3}
\eea
where the indices run over the following ranges:
$$
{\widehat J}_a^b : ( 1\leq a \leq N-1, a\leq b \leq N) \quad , \quad
\widehat{\overline G}{}^i : (1\leq i \leq N) \quad .
$$
In this non-primary basis the currents $\widehat{\overline G}{}^i $ have the
same conformal weights $3,4,...,N+2$ as the currents of $W_{N+2}$ algebra,
so the stress tensor ${\widehat {\cal T}}$ coincides with the stress tensor
of $W_{N+2}$ algebra.

It is also instructive to rewrite the $W_{N+2}^{lin}$ algebra \p{linal3} in
the primary basis  $\left\{ {\cal T},{\widehat U},
{\widehat J}{}_a^b,\widehat{\overline G}{}^i\right\}$, where a new
stress-tensor ${\cal T}$ is defined as
\be
{\cal T} = {\widehat {\cal T}}-\frac{(N+2)(K-1)}{2K} {\widehat U}{}'+
  \frac{K-1}{K}\sum_{m=1}^{N-1}  m \left( {\widehat J}{}_{N-m}^{N-m}
         \right)'
\ee
and the OPEs have the following form
\bea
{\cal T}(z_1) {\cal T}(z_2) & = &
 \frac{N+1-6\frac{(K-1)^2}{K}}{2z_{12}^4}+\frac{2{\cal T}}{z_{12}^2}+
                   \frac{ {\cal T}'}{z_{12}} \quad , \quad
{\widehat U}(z_1){\widehat U}(z_2)  =
  \left(\frac{2NK}{2+N}\right)\frac{1}{z_{12}^2} \; , \nonumber \\
{\cal T}(z_1){\widehat J}_a^b(z_2) & = &
        \frac{\left( 1-\frac{a-b}{K}\right) {\widehat J}_a^b}{z_{12}^2}+
                   \frac{{\widehat J}_a^b{}'}{z_{12}} \; ,
                      \nonumber \\
{\cal T}(z_1){\widehat U}(z_2)  & = &
    \frac{{\widehat U}}{z_{12}^2}+\frac{{\widehat U}'}{z_{12}} \;,
               \nonumber \\
{\cal T}(z_1)\widehat{\overline G}{}^i(z_2) & = &
            \frac{\left(\frac{3}{2}+\frac{1+2i}{2K}\right)
      \widehat{\overline G}{}^i}{z_{12}^2}+
                \frac{\widehat{\overline G}{}^i{}'}{z_{12}}\; , \nonumber \\
{\widehat J}_a^b(z_1){\widehat J}_c^d(z_2) & = &
  K\frac{\delta_a^d\delta_c^b-
       \frac{1}{N}\delta_a^b\delta_c^d}{z_{12}^2}+
        \frac{\delta_c^b {\widehat J}_a^d-
          \delta_a^d {\widehat J}_c^b}{z_{12}} \; , \nonumber \\
{\widehat U}(z_1)\widehat{\overline G}{}^i(z_2)  & =  &
     -\frac{\widehat{\overline G}{}^i}{z_{12}}  \; , \;
{\widehat J}_a^b(z_1)\widehat{\overline G}{}^i(z_2)  =
       \frac{-\delta_a^i \widehat{\overline G}{}^b +
 \frac{1}{N}\delta_a^b \widehat{\overline G}{}^i}{z_{12}} \; , \nonumber \\
\widehat{\overline G}{}^i(z_1) \widehat{\overline G}{}^j(z_2)  & = &
  \mbox{regular} \;. \label{linal4}
\eea
In this basis the "chain" structure of linearizing algebras $W_N^{lin}$,
i.e. the property of linearizing algebras with a given $N$ to be subalgebras
of those with a higher $N$, becomes most transparent. Namely, if we redefine
the currents of $W_{N+2}^{lin}$ as
\bea
{\cal U}_1 & = & {\widehat U}-N\sum_{m=1}^{N-1}{\widehat J}_m^m \;, \nonumber
\\
{\cal U} & = & \frac{(N+2)(N-1)}{N(N+1)}{\widehat U}+
                 \frac{2}{N+1}\sum_{m=1}^{N-1}{\widehat J}_m^m \;
                                          \nonumber \\
{\widetilde {\cal T}} & = & {\cal T}+\sqrt{\frac{N+2}{12KN^2(N+1)}}{\cal U}_1'
\; ,
 \quad \left( \mbox{or} \quad
{\widetilde {\cal T}} =  {\cal T}-
\frac{N+2}{2KN^2(N+1)}({\cal U}_1{\cal U}_1)\right) \; ,
\nonumber \\
{\cal J}_a^b & = & {\widehat J}_a^b-
            \frac{\delta_a^b}{N-1}
             \sum_{m=1}^{N-1}{\widehat J}_m^m \;,
         (1\leq a \leq N-2, a\leq b \leq N-1 ) \; ,\nonumber \\
{\cal S}_a & = & {\widehat J}_a^N \;, (1\leq a \leq N-1) \; ,\nonumber \\
{\overline{\cal G}}{}^i & = & {\widehat{\overline G}}{}^i
             \;, (1\leq i \leq N-1)\; , \nonumber \\
{\overline{\cal Q}} & = & {\widehat{\overline G}}{}^N  \; , \label{linal5}
\eea
then the subset
${\widetilde {\cal T}},{\cal U},{\cal J}_a^b,{\overline{\cal G}}{}^i $
generates
the algebra $W_{N+1}^{lin}$ in the form \p{linal4}. Thus, the
$W_{N+2}^{lin}$ algebras constructed have the following structure
\be
W_{N+2}^{lin}=\left\{ W_{N+1}^{lin}, {\cal U}_1,{\cal S}_a,
        {\overline{\cal Q}} \right\}
\ee
and therefore there exists the following chain of embeddings
\be
\ldots W_{N}^{lin} \subset W_{N+1}^{lin} \subset  W_{N+2}^{lin} \ldots \quad .
       \label{chain}
\ee
Let us stress that the nonlinear $W_{N+2}$ algebras do not possess the
chain structure like \p{chain}, this property is inherent only to their
linearizing algebras $W_{N+2}^{lin}$.

By this we finished the construction of conformal linearizing algebras
$W_{N+2}^{lin}$ which contain $W_{N+2}$ as subalgebras in a nonlinear basis.
The explicit expression for the transformations from the currents of
$W_{N+2}^{lin}$  algebra to those forming $W_{N+2}$ algebra
is a matter of straightforward calculation once we know the exact
structure of the linearizing algebra. In the next Subsections for the
particular cases of $W_3$ and $W_4$ algebras we will consider this question
in more detail.

Finally, let us stress that knowing the structure of the linearized
algebras $W_{N+2}^{lin}$  helps us to reveal some interesting properties of
the $W_{N+2}$ algebras and their representations.

First of all, each realization of $W_{N+2}^{lin}$ algebra gives rise to a
realization of $W_{N+2}$. Hence, the relation between linear and nonlinear
algebras opens a way to find new non-standard realizations of $W_{N+2}$
algebras. As was shown in \cite{BO,LPSX} for the particular case of $W_3$,
these new realizations (for details, see next Subsection) can be useful for
constructions of new string theories and solving the problem of embedding
Virasoro string into the $W_3$ one.

Among many interesting realizations of $W_{N+2}^{lin}$ there is one very
simple particular realization which can be described as follows.
A careful inspection of the OPEs \p{linal4} shows that the currents
\be
{\widehat{\overline G}}{}^i \; , \;
{\widehat J}_a^b : ( 1\leq a \leq N-1, a < b \leq N)
\ee
are null fields and so they can be consistently put equal to zero.
In this case the algebra $W_{N+2}^{lin}$ will contain only Virasoro
stress tensor ${\cal T}$ and $N$ $u(1)$-currents $\left\{ {\widehat U},
{\widehat J}_1^1 , \ldots {\widehat J}_{N-1}^{N-1}\right\}$. Of course,
there exists the new decoupling basis, where all these currents commute
with each other (see similar discussion at the end of Section 2).
One can check that the currents of $W_{N+2}$ algebra are realized in this
basis in terms of some stress tensor $T_{Vir}$ with the same central charge
$c_{Vir}$ as is given in eq. \p{vir} and $N$ decoupled commuting $u(1)$
currents. Surprisingly, the values of $c_{Vir}$ corresponding to the
minimal models of Virasoro algebra \p{level} \cite{min} induce the central
charge $c_{W_{N+2}}^{min.mod.}(p,q)$ of the minimal models for
$W_{N+2}$ algebra \cite{FL}
\be
c_{W_{N+2}}^{min.mod.}(p,q)=(N+1)\left(1-(N+2)(N+3)
\frac{(p-q)^2}{pq}\right) \label{minmodW_N}
\ee
(let us remind that the stress tensor of $W_{N+2}$ coincides with
the stress tensor $\widehat {\cal T}$ in the non-primary basis \p{linal3}).
For the $W_3$ algebra this property has been first discussed in \cite{KS}.

\subsection{Linearizing $W_3$ algebra.}

In this Subsection, as an example of our construction, we
present the explicit formulas concerning the conformal linearization
of $W_3$ algebra \cite{KS}.

The $W_3$ algebra \cite{W3} contains the currents
$\left\{{\widehat {\cal T}},{\cal W}\right\}$
with spins $\left\{ 2,3\right\}$, respectively.

The structure of the linearizing algebra $W_{3}^{lin}$ in the primary basis
can be read off from the OPEs \p{linal4} by putting $N=1$.
So, the algebra $W_{3}^{lin}$ contains the currents
$\left\{{\cal T}, {\widehat U}, {\widehat{\overline G}{}^1}\right\}$,
with the
conformal weights $\left\{ 2,1,\frac{3(K+1)}{2K}\right\}$, respectively.

Passing to the currents of $W_3$ goes over two steps.

Firstly, we must write down most general, nonlinear in the currents of
$W_3^{lin}$, {\it invertible} expressions for the currents
${\widehat {\cal T}},{\cal W}$ with the desired conformal weights
2 and 3. It can
be easily done in the nonprimary basis \p{linal3}, where the stress tensor
$\widehat {\cal T}$  coincides with the stress tensor of $W_3$ algebra.

Secondly, we should calculate the OPEs between the constructed expressions
and demand them to form a closed set.

This procedure completely fixes all the coefficients in the expressions for
the currents of $W_3$ algebra in the primary basis in
terms of currents of $W_3^{lin}$ (up to unessential rescalings).
Let us stress that we do not need to
know the explicit structure of $W_3$ algebra. By performing the
second step, we automatically reconstruct the $W_3$ algebra.

Let us present here the results of our calculations for the $W_3$ algebra
\bea
{\widehat {\cal T}} & = & {\cal T} +
\frac{3(K-1)}{2K}{\widehat U}' \; , \nonumber \\
{\cal W} & = & {\widehat {\overline G}}{}^1 +
\frac{6^{1/2}}{((5K-3)(5-3K))^{1/2}} (- ({\widehat U} {\cal T}) +
\frac{1}{K} ({\widehat U} {\widehat U} {\widehat U}) +   \nonumber \\
& & \frac{3(K-1)}{2K} ( {\widehat U} {\widehat U}' ) -
\frac{K-1}{2} {\cal T}' +
\frac{(K-1)^2}{4K}( {\widehat U}'') ) \; . \label{realizW3}
\eea

Thus, all the remarkable nonlinear features of $W_3$ algebra can be traced
to the choice of a nonlinear basis in the linear algebra $W_3^{lin}$.
For example, every realization of $W_3^{lin}$ is a realization of the $W_3$
algebra simultaneously\footnote{Of course, the inverse statement
is not correct in general.}.
So the problem of the construction $W_3$-realizations is reduced to much
more simple problem of constructing realizations of $W_3^{lin}$. In the
rest of this Subsection we present an example of such a realizations.

    From the simple structure of the $W_3^{lin}$ algebra OPEs \p{linal4} at
$N=1$ it is evident that its most general realization includes at least
two free bosonic scalar fields $\phi_i$ ($i=1, 2$) with OPEs
\be
\phi_i(z_1)\phi_j(z_2) = -\delta_{ij}ln(z) \; ,
\ee
as well as a commuting with them Virasoro stress tensor $T_r$ having
a nonzero central charge which we denote $c_{T_r}$. Representing the
bosonic primary field ${\widehat {\overline G}}{}^1$ in the standard way
by an exponential of $\phi_i$, the current ${\widehat U}$ by the
derivative of $\phi_i$
and ${\cal T}$ by the sum of $T_r$ and the standard stress-tensors of $\phi_i$
with background charges, and requiring them to satisfy the OPEs \p{linal4},
we find the following expressions
\bea
{\widehat {\overline G}}{}^1 & = &
s \cdot \mbox{exp}\left(i\sqrt{n-\frac{3}{2K}}\phi_2+
  \frac{i{\sqrt{3}}}{\sqrt{2K}}\phi_1\right)  \quad ,\nn \\
{\widehat U} & = & - i\sqrt{{\frac{2K}{3}}}\phi_1'\quad , \nn \\
{\cal T} & = & T_r-\frac{1}{2}\left( \phi_1'\right)^2  -
      -\frac{1}{2}\left( \phi_2'\right)^2  -
 \frac{i\left(3-n+\frac{3}{2K}\right)}{2\sqrt{n-\frac{3}{2K}}}\phi_2''
             \quad , \nn \\
c_{T_r} & = & 3\frac{\left( 3-n+\frac{3}{K}\right)^2}{n-\frac{3}{2K}}-
      \frac{6(K-1)^2}{K} \quad , \label{ff}
\eea
where $n \in Z_{+}$ and $s$ is an arbitrary parameter.
Its arbitrariness reflects the
invariance of the OPEs \p{linal4} with respect to rescaling of the null
field ${\widehat {\overline G}}{}^1$. If $s\neq 0$, it can always be chosen,
e. g., equal to unity by a constant shift of the field $\phi_1$.

In the case of $s=0$ the obtained realizations can be simplified by
introducing a new Virasoro stress-tensor ${\widetilde {T_r}}$ with the
central charge $c_{{\widetilde {T_r}}}$, which absorbs
the field $\phi_2$
\bea
{\widetilde {T_r}} & = & T_r-\frac{1}{2}\left( \phi_2'\right)^2  -
 \frac{i\left(3-n+\frac{3}{K}\right)}{2\sqrt{n-\frac{3}{2K}}}\phi_2''
     \quad , \\
c_{{\widetilde {T_r}}} & = & 1 - 6 \frac{(K-1)^2}{K} \label{cc}
\eea
and $\phi_2-$dependence disappears altogether. In this notation the
expressions \p{ff} are given by
\begin{eqnarray}
{\widehat {\overline G}}{}^1 & = & 0 \quad , \nn \\
{\widehat U} & = & - i\sqrt{{\frac{2K}{3}}}\phi_1' \quad , \nn \\
{\cal T} & = & {\widetilde {T_r}} -
\frac{1}{2}\left( \phi_1' \right)^2 \quad . \label{ff1}
\end{eqnarray}

After substituting eqs. \p{ff} into \p{realizW3}, we get a realization of the
$W_3$ algebra which generalizes the realization obtained in \cite{Rom1} and
is reduced to it at $s=0$.

The discussed here realizations could be a starting point for constructing
new versions of $W_3$-string theories \cite{BO,LPSX}. For example, as was
shown recently in \cite{LPSX} a special case of these realizations
corresponding to $n=0$, provides embedding of
the Virasoro string into non-critical and critical $W_3$ strings.

\subsection{Linearizing $W_4$ algebra.}

In this Subsection, we continue our consideration of algebras belonging to
$W_N$ series, and discuss the conformal linearization of $W_4$ algebra
\cite{KSlin1,KSlin2,MR}.

The $W_4$ algebra \cite{Nahm} contains the currents
$\left\{{\widehat {\cal T}},{\cal W}, {\cal V}\right\}$
with spins $\left\{ 2,3,4\right\}$, respectively.

The structure of the linearizing algebra $W_{4}^{lin}$ in the primary basis
can be read off from the OPEs \p{linal4} by putting $N=2$.
So, the algebra $W_{4}^{lin}$ contains the currents
$\left\{ {\cal T},{\widehat U},{\widehat J}{}_1^1,{\widehat J}{}_1^2,
\widehat{\overline G}{}^1,\widehat{\overline G}{}^2\right\}$, with the
conformal weights $\left\{ 2,1,1,\frac{K+1}{K},\frac{3(K+1)}{2K},
\frac{3K+5}{2K}\right\}$, respectively.

Using exactly the same arguments as given in previous Subsection,
we  write down most general, nonlinear in the currents of
$W_4^{lin}$, {\it invertible} expressions for the currents
${\widehat {\cal T}},{\cal W}, {\cal V}$ with the desired conformal
weights 2, 3 and 4. It can be easily done in the nonprimary basis
\p{linal3}, where the stress tensor $\widehat {\cal T}$  coincides
with the stress tensor of $W_4$ algebra. After that we calculate
the OPEs between the constructed expressions and demand them to form
a closed set.

The results of our calculations for the $W_4$ algebra look as follows
\bea
{\widehat {\cal T}} & = & {\cal T} +\frac{2(K-1)}{K}{\widehat U}'-\frac{K-1}{K}
       {\widehat J}{}_1^1{}' \; , \nonumber \\
{\cal W} & = & {\widehat{\overline G}}{}^1 +\frac{K-1}{K}(T_1-T_2)'+
    \frac{1}{K}\left((T_1-T_2){\widehat U}\right)-
    \frac{K-1}{K}{\widehat J}{}_1^2{}'-
    \frac{1}{K}({\widehat J}{}_1^2{\widehat U}) \; , \nonumber \\
{\cal V} & = & -{\widehat{\overline G}}{}^2 +
    \frac{K-1}{K}{\widehat{\overline G}}{}^1{}'+\frac{1}{2K}\left(
    ({\widehat J}{}_1^2{\widehat J}{}_1^2)+{\widehat J}{}_1^2{}''\right)+
    \frac{1}{K}\left( ({\widehat U}-2{\widehat J}{}_1^1)
           {\widehat{\overline G}}{}^1\right) -
        \frac{1}{K}\left( (T_1-T_2){\widehat J}{}_1^2\right) \nonumber \\
 & & + \frac{1}{2K}\left((T_1-T_2)(T_1-T_2)\right)-\frac{2}{K^2}
     \left( {\widehat J}{}_1^1{\widehat J}{}_1^2\right)'+
     \frac{1}{K^2}\left( (T_1+T_2)(2(K-1){\widehat U}'+
         ({\widehat U}{\widehat U}))\right)  \nonumber \\
 & & + \frac{K-1}{K^2}\left( (T_1+T_2)'{\widehat U}\right)+
     \frac{(K-1)^2}{2K^2}(T_1+T_2)''+
     \frac{(K-1)}{3K^2}{\widehat U}''' \nonumber \\
 & &  + \frac{(2-K)(2K-1)}{4K^3}({\widehat U}''{\widehat U}) +
         \frac{(K-1)^2}{K^3}({\widehat U}'{\widehat U}') +
\frac{16(3-2K)(3K-2)}{K(300-637K+300K^2)}({\cal T}{\cal T}) \nonumber \\
& & + \frac{-120K^4+430K^3-617K^2+430K-120}{4K^2(300-637K+300K^2)}
{\cal T}'' \; , \label{realiz}
\eea
where the auxiliary currents $T_1$ and $T_2$ are defined as
\bea
T_1 & = & {\cal T}-\frac{1}{K}({\widehat J}{}_1^1{\widehat J}{}_1^1)-
    \frac{1}{2K}({\widehat U}{\widehat U}) \; , \nonumber \\
T_2 & = & \frac{1}{K}({\widehat J}{}_1^1{\widehat J}{}_1^1)-
         \frac{K-1}{K}{\widehat J}{}_1^1{}' \;.
\eea
Let us repeat again that from the beginning we do not need to know the
explicit structure of $W_4$ algebra except for the spin content of its
currents.
By performing the calculations we automatically reconstruct the $W_4$ algebra.

For the $W_4^{lin}$ algebra \p{linal4}
the currents ${\widehat{\overline G}}{}^1,{\widehat{\overline G}}{}^2 $ and
${\widehat J}{}_1^2$ are null-fields. So we can consistently put them
equal to zero. In this case the expressions \p{realiz} provide us with the
Miura realization of $W_4$ algebra in terms of two currents with conformal
spins 2 ($T_1, T_2$) and with the same central charges,
and one current with spin 1 (${\widehat U}$) which commute with each other.

\section{Linearizing $W(sl(N+2),sl(3))$ algebras.}

In this Section following Refs. \cite{KSlin2,BKSnull2} we prove the
point ii) of the main {\it Conjecture} of
Section 3 concerning linearizing algebras for $W(sl(N+2),sl(3))$ by
explicit construction of invertible expressions for the currents
of $W(sl(N+2),sl(3))$ algebra in terms of currents of
$\widetilde{QSCA}{}_N^{lin}$ \p{linal2}.

Let us briefly remind the conformal spin content of $W(sl(N+2),sl(3))$
algebra \cite{BTD}. It consists of the following currents: spin-two stress
tensor ${\cal T}$, spin-one $sl(N-1)$ ($N>1$) and $u(1)$ affine currents
${\cal J}{}_a^b$ and ${\cal U}$, respectively, commuting with them
spin-three current ${\cal W}$, and two multiplets of spin-two currents
${\cal T}_a$ and ${\overline {\cal T}{}^a}$ having opposite $u(1)$ charge and
belonging to the fundamental and its conjugated representations of $sl(N-1)$.

Using exactly the same approach as given in Subsections 4.1 and 4.2 for the
cases of $W_3$ and $W_4$ algebras, we introduce nonprimary basis for the
currents of $\widetilde{QSCA}{}_N^{lin}$ algebra \p{linal2} with new stress
tensor
\bea
{\widehat {\cal T}} & = & {\widetilde T}  +
        \frac{(N+2)(K-N)}{2NK}{\widetilde U}' +
        {\widetilde J}{}_1^1{}' \; , \label{newstress}
\eea
with the following central charge $c_{W(sl(N+2),sl(3))}$
\bea
c_{W(sl(N+2),sl(3))}=
N^2+24N+25-24K-\frac{N^3+6N^2+11N+6}{K}. \label{newcharge}
\eea
In this basis the
currents $\left\{ {\widehat {\cal T}}, {\widetilde U}, {\widetilde J}{}_1^1,
{\widetilde J}{}_a^b, {\widetilde J}{}_a^1, {\widetilde J}{}_1^a,
\widetilde{\overline G}{}^1, \widetilde{\overline G}{}^a\right\}$
(where here $2\leq a,b \leq N$) have the spins
$\left\{ 2,1,1,1,2,0,3,2\right\}$, respectively, so stress tensor
${\widehat {\cal T}}$  coincides with the stress tensor of the
$W(sl(N+2),sl(3))$ algebra.
In this basis we write down most general, nonlinear in the currents of
$\widetilde{QSCA}{}_N^{lin}$ algebra, {\it invertible} expressions for the
currents $\left\{{\cal W}, {\cal J}{}_a^b, {\cal U}, {\cal T}_a,
{\overline {\cal T}{}^a}\right\}$ with the desired conformal weights
$\left\{ 3,1,1,2,2\right\}$, calculate the OPEs between the constructed
expressions and demand them to form a closed set.

Let us present here the results of our calculations
\bea
{\cal J}{}_a^b & = & {\widetilde J}{}_a^b + \frac{1}{N-1}
         {\widetilde J}{}_1^1\delta_a^b \; , \nonumber \\
{\cal U} & = & -\frac{2(N-1)}{N}{\widetilde U} + 2 {\widetilde J}{}_1^1
\; , \nonumber \\
{\cal T}_a & = & {\widetilde J}{}_a^1 \; , \nonumber \\
{\overline {\cal T}{}^a} & = & \widetilde{\overline G}{}^a +
    \frac{1}{3K(K-N)}
    (-\frac{2K^2-3K+2}{2}  {\widetilde J}{}_1^a{}'' -
    (K-N-1) ({\cal J}{}_b^a{}' {\widetilde J}{}_1^b) -
    (2K-1) ({\cal J}{}_b^a {\widetilde J}{}_1^b{}') \nonumber \\
& & - \frac{(N+2)(K-1)}{2N} ({\widetilde U}{}' {\widetilde J}{}_1^a) -
   \frac{(N+2)(K-1)}{N} ({\widetilde U} {\widetilde J}{}_1^a{}')+
    \frac{2K-N^2-N-2}{2(N-1)} ({\widetilde J}{}_1^1{}'
{\widetilde J}{}_1^a) \nonumber \\
& & + \frac{2K+N-2}{N-1} ({\widetilde J}{}_1^1 {\widetilde J}{}_1^a{}') -
\frac{1}{2} ({\cal J}{}_b^c {\cal J}{}_c^b {\widetilde J}{}_1^a) -
    ({\cal J}{}_b^c {\cal J}{}_c^a {\widetilde J}{}_1^b) -
\frac{N+2}{N} ({\widetilde U} {\cal J}{}_b^a
{\widetilde J}{}_1^b) \nonumber \\
& & + \frac{2}{N-1} ({\cal J}{}_b^a {\widetilde J}{}_1^1 {\widetilde J}{}_1^b)-
\frac{N^2-N+2}{2(N-1)^2} ({\widetilde J}{}_1^1 {\widetilde J}{}_1^1
{\widetilde J}{}_1^a) + \frac{N+2}{N(N-1)} ({\widetilde U}
{\widetilde J}{}_1^1 {\widetilde J}{}_1^a)  \nonumber \\
& & - \frac{(N+1)(N+2)}{2N^2} ({\widetilde U} {\widetilde U}
{\widetilde J}{}_1^a) +  K ({\widetilde T} {\widetilde J}{}_1^a) -
({\widetilde J}{}_b^1 {\widetilde J}{}_1^a
{\widetilde J}{}_1^b)),  \label{newalg1}
\eea
\bea
{\cal W} & = & -{\widetilde {\overline G}}{}^1 + \frac{1}{3(N-3K+2)}
( \frac{N+2}{N}({\widetilde T} {\widetilde U}) +
\frac{2K}{(K-N)}({\widetilde T} {\widetilde J}{}_1^1) -
\frac{N-3K+2}{2} {\widetilde T}' \nonumber \\
& & + \frac{(N+2)^2(K+5KN-N^2-3N-2)}{6KN^3(N-3K+2)}
({\widetilde U} {\widetilde U} {\widetilde U}) +
 \frac{(N+2)(K+2KN-N-2)}{N^2(K-N)(N-3K+2)}
({\widetilde U} {\widetilde U} {\widetilde J}{}_1^1)   \nonumber \\
& & - \frac{(N+2)(K-N-2)(2K-5KN+N^2+2N)}{2NK(K-N)(N-1)(N-3K+2)}
( {\widetilde U} {\widetilde J}{}_1^1 {\widetilde J}{}_1^1)  \nonumber \\
& & + \frac{(K^2(11N^2-13N+2)-K(6N^3+N^2-20N+4)+N^4+2N^3-4N^2-8N)}
{3K(K-N)(N-1)^2(N-3K+2)}
( {\widetilde J}{}_1^1 {\widetilde J}{}_1^1 {\widetilde J}{}_1^1) \nonumber \\
& & + \frac{(N-2)(N-3K+2)}{K(N-1)(K-N)}
( {\widetilde J}{}_1^1 {\widetilde J}{}_a^1 {\widetilde J}{}_1^a) +
\frac{(N+2)(N-3K+2)}{KN(K-N)}
( {\widetilde U} {\widetilde J}{}_a^1 {\widetilde J}{}_1^a)   \nonumber \\
& & - \frac{(N+2)(K-N-2)}{2KN(K-N)}
( {\widetilde U} {\cal J}{}_a^b {\cal J}{}_b^a) -
\frac{KN-K+N+2}{K(N-1)(K-N)}
( {\cal J}{}_a^b {\cal J}{}_b^a {\widetilde J}{}_1^1)  \nonumber \\
& & + \frac{(N-3K+2)}{K(K-N)}
( {\cal J}{}_a^b {\widetilde J}{}_b^1 {\widetilde J}{}_1^a) -
\frac{N-3K+2}{3K(K-N)(K-1)}
( {\cal J}{}_a^b {\cal J}{}_b^c {\cal J}{}_c^a)   \nonumber \\
& & - \frac{(N+2)(K^2+4KN-2K-N^2-2N)}{2NK(K-N)}
( {\widetilde U} {{\widetilde J}{}_1^1}{}') -
\frac{(N+2)(K+N-2)}{2N(K-N)}( {\widetilde U}' {\widetilde J}{}_1^1) \nonumber
\\
& & + \frac{N-3K+2}{K}
( {{\widetilde J}{}_a^1}{}' {\widetilde J}{}_1^a) -
\frac{K^2(5N-2)-K(3N^2+16N-4)+4N^2+8N}{2K(K-N)(N-1)}
( {{\widetilde J}{}_1^1}{}'{\widetilde J}{}_1^1)  \nonumber \\
& & + \frac{(3K^2-3NK-12K+4N+8)(N-3K+2)}{6K(K-N)(K-1)}
( {{\cal J}{}_a^b}{}' {\cal J}{}_b^a) +
\frac{2(K-1)(N-3K+2)}{K(K-N)}
( {\widetilde J}{}_a^1 {{\widetilde J}{}_1^a}{}')   \nonumber \\
& & - \frac{(N+2)(K+2NK-N-2)}{2N^2K} ( {\widetilde U}'{\widetilde U} ) -
\frac{(N+2)(3K^2+2NK-2K-N^2+4)}{12NK}( {\widetilde U}'') \nonumber \\
& & - \frac{3K^3+K^2(17N-2)-K(9N^2+22N-4)+N^3+6N^2+8N}{6K(K-N)}
( {{\widetilde J}{}_1^1}{}'')) , \label{newalg}
\eea
where here the indices $a, b,...$ run over the following ranges:
$2\leq a,b \leq N$.

Let us remind that for the $\widetilde{(Q)SCA}{}_N^{lin}$ algebra \p{linal2}
the currents ${\widetilde{\overline G}{}^a}$ are null fields and we can
consistently put them equal to zero
(see discussion at the end of Section 2).
In this case the expressions \p{newstress}, \p{newalg1}, \p{newalg} provide
us with the Miura realization of $W(sl(N+2),sl(3))$ algebra in terms of
currents ${\widetilde T}$, ${\widetilde U}$, ${\widetilde J}{}_a^b$.
Using the arguments similar to those in the end of the Section 2
we have following conjecture for the spectrum
of central charges of $W(sl(N+2),sl(3))$ algebra minimal models
\bea
c_{W(sl(N+2),sl(3))} & = &
N^2+24N+25-\frac{24p^2+(N^3+6N^2+11N+6)q^2}{pq}. \label{newminmod}
\eea

At the end of this Section we would like to briefly discuss one more
application of formulas \p{newstress}, \p{newalg1}, \p{newalg}
to the construction of the so called modulo null fields realizations
for $W_3$ algebra. In contradistinction to ordinary realizations, for
such ones in the OPE of spin-three current ${\cal W}$ with itself,
besides the standard terms, some nonzero spin-four operator ${\cal V}$
is also present\footnote{This operator could be composite or elementary.}
\bea
{\cal W}(z_1){\cal W}(z_2) & = & standard \; terms + \frac{{\cal V}}{z_{12}} +
\frac{{\cal V}'}{2z_{12}}.  \label{nullW_3}
\eea
There is one strong restriction on this operator: its OPE with itself
must contain no central term, i.e.
\bea
< {\cal V}{\cal V} >=0 , \label{nullcond}
\eea
nevertheless in the r.h.s. of this OPE another current possessing the same
property \p{nullcond} could appears. All such currents are called
null fields and form the ideal of the algebra. Due to the last property
they can be consistently set equal to zero and on the shell of these
constraints spin 2 and 3 currents form a realization of $W_3$ algebra.

 From the above definition it is clear that the problem of construction of the
realizations modulo null fields is very complicated one. However it can be
reduced to an easier task of construction of the ordinary realizations, but
for bigger algebras, containing more currents than $W_3$ and including the OPE
\p{nullW_3} among the full set of its OPEs. Then for some discrete values of
the central charge, null field condition \p{nullcond} for the spin-four
operator of such algebra could be satisfied and so at this particular values
of the central charge the realizations of such bigger algebra form
simultaneously a $W_3$ realization modulo null fields.

As has been shown in the beginning of this Section, the $W(sl(N+2),sl(3))$
algebras contain the currents with spins 2 and 3 and so on their basis we can
apply the above mentioned approach for construction of modulo null fields
realizations of $W_3$ algebra.

Following \cite{BKSnull1,BKSnull2} we introduce the new stress tensor
${{\cal T}_w}$
\bea
{{\cal T}}_w & = & {\widehat {\cal T}}  -
\frac{1}{2(K-1)} {{\cal J}{}_a^b} {\cal J}{}_b^a -
\frac{N+2}{8(N-1)(3K-N-2)} {\cal U} {\cal U} \; , \label{cosetstress}
\eea
with the central charge $c_{W_3}$
\bea
c_{W_3} & = & - \frac{(4K-N-2)(3K-N-3)(2K-N-1)}
{K(K-1)}, \label{newchargeW_3}
\eea
which together with the spin-three current ${\cal W}$ commutes with the
currents ${\cal U}$, ${{\cal J}{}_a^b}$ and belongs to the coset
\bea
\frac{W(sl(N+2),sl(3))}{u(1) \oplus sl(N-1)}.
\eea
Substituting \p{newstress}, \p{newalg1} into \p{cosetstress}, \p{newalg}
and using the basic OPEs \p{linal2}, one can check that OPE of spin 3
current with itself looks like \p{nullW_3} and spin-four operator\footnote{For
considered algebras spin-four operator ${\cal V}$ is very
complicated composite operator and we do not reproduce here its explicit
expression.}
becomes the null field at the following values of central charge $c_{W_3}$
and corresponding values of parameter $K$ \cite{BKSnull1,BKSnull2}
\bea
c_{W_3}=c_{W_3}^{min.mod.}(3,2)=-2  \Rightarrow
K=\frac{N+2}{2}, (N \neq 2);K=\frac{N+1}{3}, (N \neq 2);
K=\frac{N+3}{4};  \nonumber \\
c_{W_3}=c_{W_3}^{min.mod.}(5+N,2+N)=
\frac{2(N-7)(N+14)}{(N+2)(N+5)} \Rightarrow
K=\frac{N+5}{3}, (N \neq 7); \label{spectrum1}
\eea
\bea
c_{W_3}=c_{W_3}^{min.mod.}(4-N,6)=\frac{2(N+4)(2N+1)}{N-4} \Rightarrow
K=\frac{N+2}{6}, (N \neq 4); \label{spectrum}
\eea
where $c_{W_3}^{min.mod.}(p,q)$ is defined by eq. \p{minmodW_N} at $N=1$.
So just for these values of parameter $K$ every realization of
$\widetilde{QSCA}{}_N^{lin}$ algebra \p{linal2} induces modulo
null fields realization of $W_3$ algebra with the currents ${{\cal T}}_w$
\p{cosetstress} and ${\cal W}$ \p{newalg}.
A first attempt to classify the possible algebras which allow a contraction
to $W_N$ is made in \cite{BEHHH} (see also references therein) where the
central charge spectrum \p{spectrum1} was conjectured. The central charge
spectrum \p{spectrum} have been constructed only very recently in
\cite{BKSnull2}.

\section{Conclusion.}

In this review we have described the class of linear (super)conformal
algebras with finite numbers~ of~ generating~ currents~ which~ contain~
in~ some~
nonlinear~ basis~ a~ wide~ class~ of $W$-(super)algebras, ~including
{}~$W(sl(N+2),sl(2))$,~~ $W(sl(N|2),sl(2))$~~ (~$u(N)$-superconformal~),
$W(sl(N+2),sl(3))$ as well as $W_N$ nonlinear algebras. The discussed
algebras do not exhaust all examples of conformal linearization and
we refer the interested reader to the original papers
\cite{BKS,BKSnull1} where $W(sl(3|1),sl(3))$, $W_{2,4}$ \cite{HT,Nahm},
$WB_2$ \cite{FST} and spin 5/2 \cite{W3}
(super)algebras are analyzed. We have illustrated some applications of
conformal linearization. Thus using the relations between linearizing and
nonlinear algebras we predicted the spectrum of central charges for
$u(N)$ (Q)SCA and $W(sl(N),sl(3))$ minimal models as well as constructed
large class of their realizations, including the induced modulo null fields
realizations for $W_3$ algebra.

We do not have a rigorous proof of our main {\it Conjecture} of Section 3,
but we have shown that it works for a wide class of nonlinear $W$-algebras
corresponding to the points i) and ii) of {\it Conjecture}. It is very
interesting to extend this investigation to the technically more complicated
case of algebras associated with the point iii) \footnote{For this type
algebra among the currents of linearizing algebra
(in the basis where stress tensor coincides with stress tensor of nonlinear
one) there are present currents with negative conformal spins,
and chargeless spin-zero composite operators be constructed. So the
transformation to the basis where linearizing algebra contains nonlinear
one as subalgebra, in principle could contain an infinite number of terms
including these operators.}.
The explicit construction of the linearizing algebras $W_{N+2}^{lin}$ for
$W_{N+2}$ reveals many interesting properties of these algebras:
they have a "chain"
structure (i.e. the linear algebras with a given $N$ are subalgebras of
those with a higher $N$), the central charge of the Virasoro subsector of
these linear algebras in the parametrization corresponding to the Virasoro
minimal models, while putting the null fields equal to zero, induces the
central charge for the minimal models of $W_N$, etc. These are some of the
reasons why we believe that our conjecture is true. Nevertheless, it would
be very interesting either to fully confirm the {\it Conjecture} from some
first principles or to find the restrictions on the range of its
applicability.

Method of conformal linearization for $W(sl(N+2),H)$ ($W(sl(N|2),H)$)
algebras described here admits a natural generalization to a larger class
of nonlinear superalgebras $W(sl(N,M),H)$. This work is in progress now.

We have explicitly demonstrated in the case of $W_3$, $W_4$ and
$W(sl(N+2),sl(3))$ algebras that we do not need to know beforehand the
structure relations of the nonlinear algebras, which rapidly become very
complicated with  growth of spins of the involved currents. Once we have
constructed the linearizing algebra, we could reproduce the structure of
the corresponding nonlinear one. So, one of the open questions now is
how much information about the properties of a given nonlinear algebra we
can extract from its linearizing algebra. The answer to this question could
be important for applications of linearizing algebras to $W$-strings,
integrable systems with $W$-type symmetry, etc.. \vspace{0.5cm}

\section*{Acknowledgments.}

It is a pleasure for us to thank S. Bellucci, A. Honecker, E. Ivanov and
V. Ogievetsky for many interesting and clarifying discussions.
One of us (A.S.) is also indebted to G. Zinovjev for his interest in this
work and useful discussions.

This investigation has been supported in part by the Russian Foundation of
Fundamental Research, grant 93-02-03821, and the International Science
Foundation, grant M9T000.


\begin{thebibliography}{99}
\bibitem{W3} A.B. Zamolodchikov, Theor. Math. Phys. 63 (1985) 1205.
\bibitem{BS} L. Feher, L. O'Raifeartaigh, P. Ruelle, I. Tsutsui and A. Wipf,
    Phys. Rep. 222 (1992) 1; \\
P. Bouwknegt and K. Schoutens, Phys. Rep. 223 (1993) 183.
\bibitem{KS} S. Krivonos and A. Sorin, Phys. Lett. B335 (1994) 45.
\bibitem{KSlin1} S. Krivonos and A. Sorin, ``Linearization of Nonlinear
           $W$-Algebras'', in Proc. Int. Workshop ``Finite
           Dimensional Integrable Systems'', July 18-21, JINR, Dubna, 1994.
\bibitem{BKS} S. Bellucci, S. Krivonos and A. Sorin,
             Phys. Lett. B347 (1995) 260.
\bibitem{KSlin2} S. Krivonos and A. Sorin, ``More on the Linearization of
           $W$-Algebras'', preprint JINR E2-95-151, hep-th/9503118.
\bibitem{TB} J. de Boer and T. Tjin, Commun. Math. Phys. 160 (1994) 317.
\bibitem{MR} J.O. Madsen and E. Ragoucy, ``Secondary Quantum Hamiltonian
         Reductions'', Preprint ENSLAPP-A-507-95, hep-th/9503042.
\bibitem{BO} F. Bastianelli and N. Ohta, ``Note on $W_3$ Realizations of the
             Bosonic Strings'', preprint NBI-HE-94-51, OU-HET 203,
             hep-th/9411156.
\bibitem{LPSX} H. L\"{u}, C.N. Pope, K.S. Stelle  and K.W. Xu,
	      Phys. Lett. B351 (1995) 179.
\bibitem{LPX} H. L\"{u}, C.N. Pope and K.W. Xu,
        ``Higher-spin Realisations of the Bosonic String'',
          Preprint CTP TAMU-10/95, hep-th/9503159.
\bibitem{BTD} F.A. Bais, T. Tjin and P. van Driel, Nucl. Phys. B357 (1991) 632.
\bibitem{FRS} L. Frappat, R. Ragoucy and P. Sorba,
             Comm. Math. Phys. 157 (1993) 499.
\bibitem{DFRS} F. Delduc, L. Frappat, R. Ragoucy and P. Sorba,
             Phys. Lett. B335 (1994) 151.
\bibitem{KB} V. Knizhnik, Theor. Math. Phys. 66 (1986) 68; \\
             M. Bershadsky, Phys. Lett. B174 (1986) 285.
\bibitem{Rom} L. Romans, Nucl. Phys. B357 (1991) 549.
\bibitem{F} J. Fuchs, Phys. Lett. B262 (1991) 249.
\bibitem{FL} V. Fateev and S. Lukyanov, Int. J. Mod. Phys. A3 (1988) 507;\\
             A. Bilal and J.-L. Gervais, Nucl. Phys. B314 (1989) 646;
               B318 (1989) 579.
\bibitem{Nahm} R. Blumenhagen, M. Flohr, A. Kliem, W. Nahm, A. Recknagel and
               R. Varnhagen, Nucl. Phys. B361 (1991) 255; \\
               H.G. Kausch and G.M.T. Watts, Nucl. Phys. B354 (1991) 740.
\bibitem{P} A. Polyakov, Int. J. Mod. Phys. A5 (1990) 833.
\bibitem{B} M. Bershadsky, Commun. Math. Phys. 139 (1991) 71.
\bibitem{A} M. Ademollo et al., Phys. Lett. B62 (1976) 105; \\
     Nucl. Phys. B111 (1976) 77; B114 (1976) 297.
\bibitem{min} A. Belavin, A. Polyakov and A. Zamolodchikov, Nucl. Phys.
          B241 (1984) 333.
\bibitem{BFK} W. Boucher, D. Friedan, A. Kent, Phys. Let. B172 (1986) 316.
\bibitem{Rom1} L. Romans, Nucl. Phys. B352 (1991) 829.
\bibitem{BKSnull1} S. Bellucci, S. Krivonos and A. Sorin,
             ``Null Fields Realizations of $W_3$ from $W(sl(4),sl(3))$
               and $W(sl(3|1),sl(3))$ Algebras'', preprint
               LNF-95/048 (P), JINR E2-95-376, hep-th/9509072.
\bibitem{BKSnull2} S. Bellucci, S. Krivonos and A. Sorin,
             ``Null Fields Realizations of $W_3$ from $W(sl(N),sl(3))$
               Algebras'', in preparation.
\bibitem{BEHHH} R. Blumenhagen, W. Eholzer, A. Honecker, K. Hornfeck and
	       R. Hubel, Int. J. Mod. Phys. A10 (1995) 2367.
\bibitem{HT} K.-J. Hamada and M. Takao, Phys. Lett. B209 (1988) 247;\\
             P. Bouwknegt, Phys. Lett. B207 (1988) 295;\\
             D.-H. Zhang, Phys. Lett. B232 (1989) 323.
\bibitem{FST} J.M. Figueroa-O'Farrill, S. Schrans and K. Thielemans,
             Phys. Lett. B263 (1991) 378.
\end{thebibliography}
\end{document}